\documentclass[10pt, conference, letterpaper]{IEEEtran}
\IEEEoverridecommandlockouts

\usepackage{booktabs} 
\usepackage{acronym}
\usepackage{float}
\usepackage{tikz}
\usepackage{amsmath}
\usepackage{amssymb}
\usetikzlibrary{fit,arrows}
\usepackage{siunitx, color}
\usepackage[commentsnumbered]{algorithm2e}
\usepackage[capitalise]{cleveref}
\usepackage[footnote,draft,silent,nomargin]{fixme}

\usepackage{flushend}  
\usepackage{supertabular, subfigure, url}
\usepackage[T1]{fontenc}

\begin{document}
\title{Spectral Graph Forge:\\Graph Generation Targeting Modularity\thanks
{This work was conducted when Luca Baldesi visited UC Irvine in 2016. All authors were partially supported by NSF Award III-1526736. A. Markopoulou is a member of the Center for Pervasive Communications and Computing (CPCC).}}

\author{\IEEEauthorblockN{Luca Baldesi}
\IEEEauthorblockA{DISI Dept\\
University of Trento\\
luca.baldesi@unitn.it}
\and
\IEEEauthorblockN{Carter T. Butts}
\IEEEauthorblockA{Sociology Dept\\
University of California, Irvine\\
buttsc@uci.edu}
\and
\IEEEauthorblockN{Athina Markopoulou}
\IEEEauthorblockA{EECS Dept, CPCC\\
University of California, Irvine\\
athina@uci.edu}
}

\newcommand{\val}[2]{
	\SI[parse-numbers = false]{\text{#1}}{#2}
}

\newcommand{\ctb}[1]{{\color{red}[CTB: #1]}}
\newcommand{\lucacomment}[1]{{\color{green}[LB: #1]}}

\newcommand{\ER}{Erd\H{o}s-R\'{e}nyi\xspace}
\newcommand{\BA}{Barab\'{a}si-Albert\xspace}
\newcommand{\SGF}{Spectral Graph Forge\xspace}
\newcommand{\TRAJ}{Trajanovski\xspace}

\newcommand{\stdimg}[3]{\begin{figure} \centering \includegraphics[width=1\columnwidth]{img/#1} \caption{#2} \label{#3} \end{figure}}
\newcommand{\citeappendix}{Details on this investigation are deferred to Appendix B.}
\newcommand{\citeappendixA}{(see Appendix A)}

\maketitle

\begin{abstract}
	Community structure is an important property that captures inhomogeneities common in large networks, and modularity is one of the most widely used metrics for such community structure. 
	In this paper, we introduce a principled methodology, the \SGF, for generating random graphs that preserves community structure from a real network of interest, in terms of modularity.  Our approach leverages the fact that the spectral structure of matrix representations of a graph encodes global information about community structure.  The \SGF uses a low-rank approximation of the modularity matrix to generate synthetic graphs that match a target modularity within user-selectable degree of accuracy, while allowing other aspects of structure to vary. We show that the \SGF outperforms state-of-the-art techniques in terms of accuracy in targeting the modularity and randomness of the realizations, while also preserving other local structural properties and node attributes. We discuss extensions of the \SGF to target other properties beyond modularity, and its applications to anonymization.
\end{abstract}

\section{Introduction}

Generating random graphs with certain prescribed properties is an increasingly important field~\cite{Mahadevan06,orsini2015quantifying,gjoka2015,motifs,graphlets}.
Random graphs are widely used to model structure in social, computer, and biological systems, with applications ranging from importance sampling to the generation of synthetic data sets. One use of synthetic graphs is as a proxy of real-world data sets. For example, in the measurement community, researchers collect Internet~\cite{survey_net_dump,autonomous_routers} or online social network~\cite{measure_osn,osn_user_behaviour} topologies, making them available to other researchers for use in studying their properties or evaluating the performance of graph algorithms.  In this context, it is often desirable to not release the original graph itself, but to generate synthetic graphs that (exactly or approximately) specify properties of interest of the real (or ``target'') graph, while varying with respect to other properties. This can be useful for anonymization purposes ({\em i.e.,} to obfuscate individual or other sensitive information) and/or for simulation purposes ({\em i.e.,} to ensure acceptable performance over a range of network structures, while preserving features known to be present in real-world use cases).


Prior work has primarily focused on generating synthetic graphs with certain local structural properties, such as degree distribution, degree correlations, subgraph counts etc.~\cite{Mahadevan06,gjoka2015,motifs,graphlets,hyper_graphs}, while global properties have received less attention. In this paper, we are interested in an important global property exhibited by many social and communications networks, namely {\em community structure}, which captures the complex pattern of inhomogeneities that characterize them. Real-world networks are typically irregular, containing numerous and often nested or overlapping subsets of vertices that are internally cohesive while being less well-connected to other subsets.  This \emph{community} \cite{newman2010networks}  or \emph{cohesive subgroup} structure \cite{wasserman.faust:bk:1994} has a profound impact on phenomena such as diffusion or robustness to failure; it can also affect the performance of sparse-graph algorithms for tasks such as path or cycle counting that are sensitive to the presence of small, densely connected clusters of vertices.  Graph generation methods that focus on local features (such as subgraph counts, the degree distribution,  clustering etc.) do not necessarily preserve community structure, and the resulting graphs may be poor proxies for real-world networks in applications that are sensitive to it, such as diffusion. 
 Therefore, distinct methods are needed to target it.


In this paper, we present a novel random graph generation framework that is well-suited to the above applications. In particular, we target community structure, within desired accuracy while allowing other features to vary. We use {\em modularity}, a well-known property  of matrices to capture the community structure \cite{newman2010networks}. Starting from a given real graph as input, we use the original graph's high-level structure captured by the eigenstructure of its modularity matrix  to derive a class of random graphs sharing the  same modularity value, within a level of accuracy.


Our approach can be summarized as follows and is depicted in \cref{fig:pipeline}.  
 Given the adjacency matrix of a graph $A$, we capture its community structure via its modularity matrix ($B$). Because $B$ is symmetric and real, it admits a spectral decomposition~\cite{axler1997linear} which expresses the matrix in terms of a set of eigenvectors and eigenvalues.  We apply a low-rank spectral approximation~\cite{lowrankapprox} on this eigenstructure to yield an approximated matrix reflecting the features most strongly associated with global structure; this matrix is then transformed and normalized to produce an edge probability matrix ($A^{\dagger}$), which is then passed to a sampling algorithm to yield a synthetic graph (with adjacency matrix $A'$).  The synthetic graphs produced will preserve the higher-order properties (modularity) of the input, while allowing other details to vary.  
 
 \SGF has several strengths. First, our approach is principled, based on spectral approximation of matrices, thus the name \SGF (SGF), and provides a tunable parameter to tradeoff between accuracy of the targeted modularity and the randomness of the produced realizations. Second, \SGF outperforms state-of-the-art baseline approaches in several aspects and for a range of graphs: (i)  it achieves the target modularity more accurately than baselines  (\cref{sec:modres})  (ii) it preserves some metrics of local structure, beyond modularity (\cref{sec:other_metrics})  (iii)  it can maintain the node and edge attributes, hence it can preserve properties beyond just pure topology (\cref{sec:attributes}). Third,  the analysis of the entropy of graphs produced  shows  that they vary substantially on dimensions other than those deliberately targeted  (\cref{sec:deanonym}); we discuss how to exploit this property and use SGF as anonymization technique for social networks. Last but not least, SGF is not limited to targeting modularity, which is the focus of this paper. The general SGF framework can be used  to target other matrix representations of a graph, such as the adjacency matrix or one of its derived matrices (including but not limited to modularity, laplacian, clique co-membership).

The structure of the rest of the paper is as follows. In \cref{sec:relwork}, we review related work. In \cref{sec:modularity}, we discuss community structure and its relation to modularity. In \cref{sec:gsf}, we present our  \textit{\SGF} Framework, and how to apply it specifically to generate graphs with a prescribed modularity. In \cref{sec:dataset}, we evaluate our algorithms for various datasets (both synthetic  and real-world social networks), against state-of-the art algorithms targeting modularity. 
\cref{sec:conclusion} concludes the paper.



\section{Related Work}
\label{sec:relwork}
There is a significant amount of work on generation of graphs that preserve {\em local} structural properties (e.g., degree distribution, small subgraph counts).  Graphlets~\cite{graphlets}, motifs~\cite{motifs} and dK-series~\cite{Mahadevan06, gjoka2015} graph generators are all intended to produce graphs that preserve such properties, starting from an initial realization,  or from a vector of input statistics. Exponential family random graph models (ERGMs) \cite{ergms} can be employed to simulate graphs having the same {\em expected} values for a series of targeted features as the original, and most applications to date have likewise focused on local properties (including properties that depend on nodal or edgewise covariates); among other differences from the above, ERGMs preserve only expected statistics (not exact values), but provide maximum entropy guarantees.  

In contrast, for global properties such as the graph modularity~\cite{newman2010networks},  there exist only a few approaches so far that target them by design.  \TRAJ et al.~\cite{Trajanovski13}~
propose an algorithm for generating graphs with a prescribed modularity. The Karrer and Newman stochastic block model~\cite{dcsbm} 
 generates graphs with a given community structure and modularity value.  In principle, other latent structure models such as the latent space models \cite{hoff:jasa:2002} and latent stochastic block models \cite{snijders.nowicki} could also be used to simulate graphs with related properties, although applications to date have been focused on inference.
 In \cref{sec:baselines}, we describe the modularity-targeting algorithms by \TRAJ et al. and Karrer and Newman, and we used them as baselines for comparison.
The relation between modularity and eigenstructure has been investigated mainly for applications such as community detection~\cite{newman2010networks,spectral_modularity}, visualization \cite{seary.richards:ch:2003}, and analysis \cite{hoff:jasa:2002}. To the best of our knowledge, our work is the first to design and implement a general framework for generation of synthetic graphs with specified high-level structure (modularity being the focus in this paper) by means of spectrally transformed inputs.

\section{Modularity and Community Structure}
\label{sec:modularity}

\subsection{Definitions and Intuition}


\emph{Modularity} is one of the most widely used metrics for characterizing community structure \cite{PNAS_2006_Newm}. It expresses the extent to which a partition of vertices divides the graph such that within-group densities are higher than between-group densities; the maximum modularity obtainable over all partitions is hence a global measure of the extent to which a graph approximates a union of dense structures that are minimally connected to one another.  Modularity is closely related to \emph{cohesive subgroup} structure, a topic of historical sociological interest~\cite{coleman1964introduction,freeman2004development}, and has been the object of extensive study within the large literature on \emph{community detection}~\cite{PNAS_2006_Newm,newman2010networks,spectral_modularity}.  

Formally, modularity is defined as follows.  Given a graph $G=(V,E)$, where $|V|=n$, with adjacency matrix $A$ and degree vector $K=(k_1,\dots,k_n)$, the modularity matrix $B$ is defined as \cite{newman2010networks}
\begin{equation}
B_{ij}= A_{i,j} - \frac{k_ik_j}{|K|}
\end{equation}
where $|K|=\sum_{i=1}^n k_i$ is the degree sum. 
 $B$ effectively captures the number of edges within nodes of the same community compared to a random graph with the same expected degree sequence. Given a node partition $\{C_1,\dots,C_m\}$ with nodes assigned to communties $c_1,\dots,c_n\in \{C_1,\dots,C_m\}$, the modularity value measuring this concentration over all node pairs is:
\begin{equation}
\label{eq:modularity} 
 Q=\frac{1}{|K|}\sum_{i,j}\left( A_{i,j} - \frac{k_ik_j}{|K|}\right) \delta(c_i,c_j).
\end{equation}
where $\delta(a,b)=1$ iff $a=b$ and 0 otherwise.
 
The notion of modularity is used in three ways. First, given a partition of nodes into communities $c_1,\dots,c_n$, $Q$ provides a metric  for how  well the partitioning matches the actual topology. 
Second, $Q$ can be used to search for community structure, by seeking a partition of nodes into $m^*$ communities, $c^*_1,\dots,c^*_n$, such that $Q^*$ is maximized.  Finally, the maximized value of $Q^*$ over all partitions can be taken as an indicator of the extent to which the graph naturally divides into distinct communities.  This quantity is sometimes referred to generically as the \emph{modularity of the graph}.
%
%



\subsection{Modularity and Spectral Structure}
The eigenstructure of an adjacency matrix $A$ associated to a given graph is intimately related to the structure of the associated graph.
$A$ can be decomposed in a set of eigenvalues $\lambda_i$ and eigenvectors $v_i$; the eigenvectors associated with positive eigenvalues describe \textit{core-periphery} structures~\cite{borgatti.everett:sn:1999} and the signs of their entries describe possible graph bi-partitions.
The eigenvectors associated with negative eigenvalues relate instead to memberships in \textit{bi-partitions}, so that, the sign of their entries indicate a likelihood of node connection.

Hence, there is an obvious affinity between the information described by the eigenvectors and modularity but for the leading eigenvector, associated to the main core of the network (the whole graph) and it would be natural to consider the eigenstructure of the residual matrix $A-\lambda_1v_1v_1^T$ (where $\lambda_1,v_1$ are respectively the leading eigenvalue and eigenvector) as a basis for community detection.

However, in practice, $v_1$ is typically correlated with $K$, and hence the modularity matrix $B=A-\left(KK^T\right)/|K|$ is very close to the residual matrix.  In keeping with this observation, it is common practice (e.g., \cite{newman2010networks}) to work with the eigenstructure of $B$ to identify partitions, recursively subdividing nodes into classes based on the signs of their eigenvector elements associated with high-modulus, positive eigenvalues of $B$. For example, in \cite{PNAS_2006_Newm}, Newman showed that the leading eigenvector of the modularity matrix  $B$ determines the best bipartition of nodes into two communities.


{\bf Beyond modularity:} While in this paper, we focus on targeting modularity, our spectral decomposition framework is more general and can be tuned to capture desired structural elements of interest, while randomizing out other features. More generally, we could also apply SGF to matrices other than $B$ (namely $A$ and its derived matrices), and use it to target other global properties.  For example, while the leading eigenvector defines the best bipartition \cite{PNAS_2006_Newm}, positive/negative eigenvalues capture core-periphery/bipartition structures, respectively.

\begin{figure}[t!]
	\begin{tikzpicture}
	\tikzstyle{every node}=[rounded corners=3pt, draw, very thick, shape=rectangle, text width=1em, node distance=1cm, text badly centered];
	\tikzset{edge/.style = {->,> = latex',line width=0.7mm}}

	\node[draw=none, text width=6em] (trasftext) {Transformation};
	\node[text width=6em, below of=trasftext, node distance=0.65cm, thin] (MA) {$M=A$};
	\node[text width=6em, below of=MA, node distance=0.8cm, fill=gray!50] (MB) {$\mathbf{M=}\mathbf{A-\frac{KK^T}{|K|}}$};
	\node[fit=(MA) (MB) (trasftext) ] (TRAN) {};
	\node[text width=6em, right of=TRAN, node distance=3.2cm, fill=gray!50] (SD) {low-rank $\alpha$-approximation of $\mathbf{M\rightarrow \tilde{M}}$};

	\node[text width=1em, shape=circle, above of=TRAN, node distance=2cm] (A) {$A$};

	\node[draw=none, text width=5em, right of=SD, node distance=3cm] (retrasftext) {Back-Transform.};
	\node[text width=5em, below of=retrasftext, node distance=0.8cm, thin] (AM) {$\tilde{A}=\tilde{M}$};
	\node[text width=5em, below of=AM, fill=gray!50] (BM) {$\mathbf{\tilde{A}=}$\\$\mathbf{\tilde{M}+\frac{KK^T}{|K|}}$};
	\node[fit=(AM) (BM) (retrasftext) ] (RETRAN) {};

	\node[draw=none, text width=7.6em, left of=BM, node distance=3cm] (sampletext) {Normalizing};
	\node[below of=sampletext, text width=7.6em, node distance=0.9cm, thin] (logit) {$A^{\dagger}=\text{logistic}(\tilde{A},k)$};
	\node[below of=logit, text width=7.6em, node distance=1cm, fill=gray!50] (truncate) {$\mathbf{A^{\dagger}=\textbf{truncate}(\tilde{A})}$};
	\node[below of=truncate, text width=7.6em, node distance=0.8cm, thin] (equalize) {$A^{\dagger}=\text{scale}(\tilde{A})$};
	\node[fit=(logit) (truncate) (equalize) (sampletext) ] (N) {};

	\node[draw=none, text width=6em, below of=TRAN, node distance=3.4cm] (bernotext) {Sampling};
	\node[below of=bernotext, text width=6em, node distance=0.9cm, fill=gray!50] (berno) {$\mathbf{A^{'}=}$\\$\mathbf{\textbf{Bernoulli}(A^{\dagger})}$};
	\node[fit=(berno) (bernotext) ] (S) {};

	\node[text width=1em, shape=circle, below of=S, node distance=1.9cm] (bA) {$A'$};

	\node[fit=(TRAN) (SD) (RETRAN) (N) (S), dashed] (process) {};
	\draw[edge] (A) -> (TRAN);
	\draw[edge] (TRAN) -> (SD);
	\draw[edge] (SD) -> (RETRAN);
	\draw[edge] (RETRAN) |- (N);
	\draw[edge] (1.6,-3) -| (S);
	\draw[edge] (S) -> (bA);


\end{tikzpicture}
	\vspace{-20pt}
	\caption{The pipeline of the \SGF (SGF) Framework. Given an undirected graph adjacency matrix $A$ as input, 
	SGF outputs a ``similar'' one $A^{'}$ from which we can build the corresponding graph, called SGF($\alpha$). Sub-blocks indicate mutually exclusive options for each step. The focus of this paper is on using SGF to target modularity, by setting $M=B=A-\left(KK^T\right)/|K|$), and the corresponding blocks are highlighted in grey.}
	\label{fig:pipeline}
	\vspace{-10pt}
\end{figure}
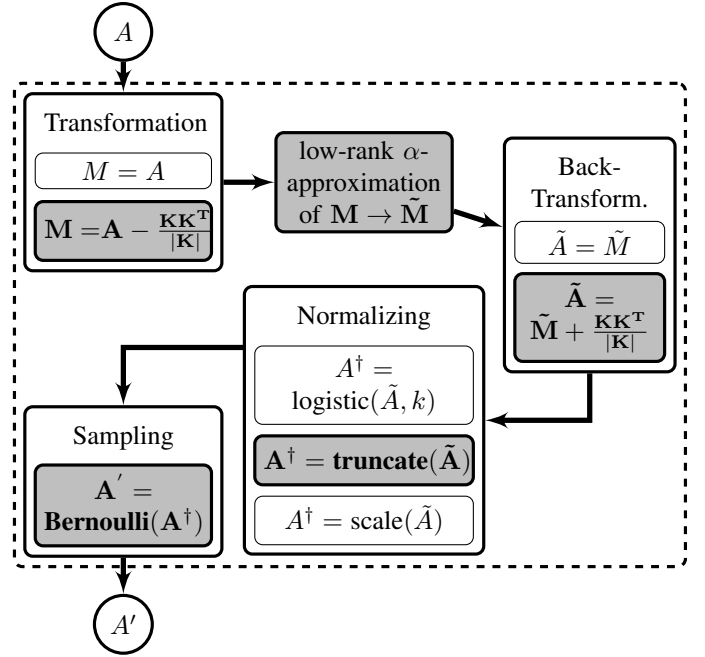

\section{\SGF}
\label{sec:gsf}
The \SGF takes an undirected graph $G$ with $n$ nodes as input and, for each run, produces a random output graph $G'$ with the same number of nodes.
We represent $G$ and $G'$ with their respective adjacency matrices $A,A'\in \{0,1\}^{n\times n}$. 
%
The \SGF is designed to be as modular and extensible as possible; its graph construction process can be viewed, hence, as a pipeline composed by several nearly independent steps, each of which can be varied as needed by the user. A representation  of this pipeline is depicted in \cref{fig:pipeline}.

\subsection{Transformation}
To allow the \SGF to preserve structure associated with either $A$ or a derived matrix, we work with a user-selected real-valued, symmetry preserving transformation of $A$, denoted $M$.  The main use case of $M$ in this paper is the modularity matrix  $M=B=A-\left(KK^T\right)/|K|$. Other examples of $M$ include the adjacency matrix itself $M=A$, and the clique co-membership matrix (whose $i,j$ cells are the counts of cliques containing both $i$ and $j$).

\subsection{Low-rank $\alpha$ approximation.} 
Since $M$ is real and symmetric, it can be expressed as
\vspace{-5pt}
\begin{equation}
	\label{eq:decomp}
	M = V\Lambda V^T = \sum_{i=1}^{n}\lambda_i v_i v_i^T,
\end{equation}	
where $\lambda_i, v_i$ are respectively the i\textit{th} eigenvalue and the i\textit{th} eigenvector of $M$ (all of which are real).
W.l.o.g., we scale $v_i$ so that $||v_i||_2=1$ and sort the eigenvectors/eigenvalues such that $|\lambda_1|\geq \dots \geq |\lambda_n|$.
Hence the first components of \cref{eq:decomp} contribute to $M$ more than the last ones.
We can perform a lossy compression of $M$ by removing the least important frequencies.
We use a parameter $\alpha\in [0,1]$ to control the fraction of the eigenvectors we want to keep from the original $M$ and our approximation is
\vspace{-5pt}
\begin{equation}
\label{eq:approx}
\tilde{M} =  \sum_{i=1}^{\lceil\alpha n\rceil} \lambda_i v_i v_i^T
\end{equation}
This is essentially a low-pass filter on $M$, which retains global characteristics captured by the first $\alpha n$ eigenvectors of the matrix, while removing the remaining ones, associated with idiosyncratic local structures.\footnote{We could also filter on phase (a \emph{graph polarizing filter}) to extract core-periphery (positive eigenvalues) or bipartite (negative eigenvalues) structures.}
The euclidean norm of the error,  $M - \tilde{M}$, due to transformation can be computed as 
\begin{equation}
\label{eq:error}
\left\lVert M - \tilde{M} \right\rVert_2 = \left\lVert \sum_{i=\lceil\alpha n\rceil+1}^{n} \lambda_i v_i v_i^T \right\rVert_2 = \lambda_{\lceil\alpha n\rceil+1}
\end{equation}
Hence, SGF can target the reproduction of $M$ within arbitrary precision, by tuning $\alpha$ \citeappendixA.

\subsection{Back-Transfomation.} 
Once the low-rank approximation $\tilde{M}$ 
 is obtained, we need to transform it to obtain an approximated adjacency matrix $\tilde{A}\in \mathbb{R}^{n\times n}$.
Where the back-transformation is the inverse of the transformation $M$ to $\tilde{M}$. 
For the modularity matrix $M=B=A-\left(KK^T\right)/|K|$, the back transformation is intuitive: $\tilde{A}=\tilde{M}+\left(KK^T\right)/|K|$. Other choices of back-transformation may be employed where $M$ is not invertible (e.g., in the case of the co-clique matrix), resulting in a different approximation.
Since $\tilde{A}$ is not, in general, an adjacency matrix, 
 our SGF fills the gap between $\tilde{A}$ and the final $A'$, by normalizing the values to be within $[0,1]$ ($ A^{\dagger}$)  and by using a stochastic process that forces the elements to be either $1$ or $0$ ($A'$), as described next.

\subsection{Normalization} \label{}
We first normalize the values of the latter in $[0,1]$ obtaining: $A^{\dagger}_{i,j} = \text{norm}(\tilde{A}_{i,j})$,
where \textit{norm} is a user-specified normalization function.
With modularity in mind, we investigated three variants for this normalization function:
\begin{itemize}
	\item \textit{logistic$(\tilde{A}_{i,j},k)$} =  $\frac{1}{1 + e^{(0.5-\tilde{A}_{i,j})\cdot k}}$, where $k\in [2,10]$
	\item \textit{truncate$(\tilde{A}_{i,j})$} =  $\left\{\begin{array}{l l}
			1 & \quad \text{if } \tilde{A}_{i,j} \geq 1 \\ 
			0 & \quad \text{if } \tilde{A}_{i,j} \leq 0 \\ 
			\tilde{A}_{i,j} & \quad \text{otherwise} \\
							\end{array} \right. $
	\item \textit{scale$(\tilde{A}_{i,j})$} =  $\frac{\tilde{A}_{i,j}-\min_{s,t}\tilde{A}_{s,t}}{\max_{s,t}\tilde{A}_{s,t}-\min_{s,t}\tilde{A}_{s,t}}$
\end{itemize}
where $k\in [2,10]$ is a parameter we use to tune the inclination of the \textit{logistic()} function.
%

%
%
We have extensively evaluated the above normalization functions and we found that the truncation rule gives the cleanest  entropy and distance relationships; in fact, \textit{scale} introduces a distortion highly dependent on the difference between the minimum and maximum value while \textit{logistic} (for which we found with $k=6$ provides the best performance for our case of study) shows poor sensitivity to $\alpha$, preserving relatively little structure throughout.
\citeappendix

\subsection{Sampling} Taking $A^{\dagger}$ as a matrix of expectations for $A'$, we complete our process by drawing $A'$ from an inhomogeneous Bernoulli graph distribution: $ A'_{i,j} = \text{bernoulli}(A^{\dagger}_{i,j}),\ \forall j>i $. Note that to generate a simple undirected graph $A'$, we set:
\begin{itemize}
	\item $A'_{i,i} = 0,\ \forall 1\leq i\leq n$
	\item $A'_{j,i} = A'_{i,j},\ \forall j>i$
\end{itemize}
As with other aspects of the pipeline, this can be generalized (e.g., to allow for degree or other constraints).
In the following, we denote by SGF$(\alpha)$ as the graph resulting from $A'$, drawn using our method on fraction $\alpha$ of the eigenvectors.

%

{\em Entropy of the Synthetic Graphs.}
One advantage of the use of an inhomogeneous Bernoulli graph in our pipeline is that we can compute exactly the (Shannon) entropy of our synthetic distribution.
Entropy indicates the degree of variation in the output graphs, and is an indicator of the extent to which a graph generation method is able to provide a range of structures compatible with its arguments; at constant graph order $n$, it necessarily varies between $n(n-1)/2$ bits for a uniform random graph and 0 bits for a generator that produces only a single realization with certainty.  Given a fixed level of preservation of targeted features, generators with higher entropy are typically preferable (as they vary the distribution of structures as widely as possible, given the desired constraints).  The ability to easily compute output entropy is a useful feature of our approach; by contrast, most other approaches provide no known way of computing output entropy, making it difficult to assess the extent to which they are able to produce a diverse range of structures.

%

Since $A'$ is drawn from a Bernoulli graph distribution with parameter matrix $A^{\dagger}$, the entropy is immediate: 
{\small
$$
H(A^{\dagger}) = \frac{1}{S} \sum_{i=1}^{n}\sum_{j=i+1}^{n} -A^{\dagger}_{i,j}\log_2[A^{\dagger}_{i,j}] -(1-A^{\dagger}_{i,j})\log_2[1-A^{\dagger}_{i,j}],
$$
}
which is readily computed from $A^{\dagger}$, where $S=\tfrac{-n(n-1)}{2}(\log_2\delta + \log_2(1-\delta))$ is the normalization factor and $\delta=\tfrac{2}{n^2-n}\sum_{i,j}A^{\dagger}_{i,j}$ is the expected graph density under $A^{\dagger}$.
This index can be interpreted as the fraction of the maximum possible entropy that $A'$ attains, given its expected density.

\subsection{Computational Complexity}
To  analyze the  complexity of  SGF, we consider the complexity of each block in \cref{fig:pipeline}. The pipeline starts with a graph $G=(V,E),~|V|=n, |E|=m$. The \textit{Transformation} and \textit{Back-Transformation} blocks can clearly be performed in $O(n^2)$. The \textit{Normalizing} and \textit{Sampling} blocks operate element-wise on {$n\times n$} matrices and apply a float operation so their cost is $O(n^2)$.
The \textit{Low-rank $\alpha$-approximation} block performs a spectral decomposition on a $n\times n$ matrix which can be done in $O(n^3)$ or $O(nm)$ in case of an input sparse matrix\footnote{Since $M$ symmetric, we can find an orthogonal matrix $Q$ such that the similarity transform $T=Q^TAQ$ gives a tridiagonal matrix~\cite{newman2010networks}, which belongs to $O(nm)$ with the Lanczos algorithm~\cite{lanczos}.
The eigenvectors $v_i=Qw_i$ of $M$ are hence directly computable from the eigenvectors $w_i$ of $Q$.
Since $Q$ is tridiagonal, its eigenvectors can be computed with the \textit{QL algorithm} with a cost of $O(n)$~\cite{newman2010networks}.}.
Overall, the bottleneck of SGF is the spectral decomposition, which, for sparse input matrices $M$, has hence a complexity of $O(n^2)$~\cite{newman2010networks}. 
We mainly target social graphs where the community structure analysis is of main interest; these graphs are generally quite sparse with a low number of nodes (each of the network in our real-world dataset has less then $10^5$ nodes).
Furthermore, SGF can gracefully tradeoff complexity for accuracy by tuning the parameter $\alpha$: keeping a few eigenvectors in the approximation (\cref{eq:approx}), increases the error (\cref{eq:error}) but decreases the complexity of the spectral decomposition (e.g. using the moment method).


\section{Evaluation}
\label{sec:dataset}

\subsection{Evaluation Setup}

To validate our approach, we test the \SGF against state-of-the art baselines (\cref{{sec:baselines}})  on several datasets (\cref{sec:testdatasets}), and in terms of several metrics.

\subsubsection{Datasets}\label{sec:testdatasets} We use both synthetic and real-world datasets; the former for a fair comparison on well-known state-of-the-art datasets in the field of community detection, and the latter to show that our results hold with real data as well.

{\bf I. Synthetic Datasets.} A random graph generator that controls modularity is the following: fix the number of communities and then control the intra-community vs. inter- community probability of edge: fixing the inter-community edge probability, higher values of intra-community edge probability leads to higher modularity value. Although stylized, this model allows for explicit control of the inherent modularity of the synthetic  graph. This idea has been widely used since the early days of community detection. For example, Girvan and Newman~\cite{girvan2002community} generated graphs with 128 nodes, communities of the same size and essentially constant node degree.  The idea has been extended by Lancichinetti et al.~\cite{Lancichinetti08} to produce networks of 1000 nodes and geometrically distributed node degrees and community sizes. We use both generators to produce ten networks from each, and refer to them as   {\bf Girvan} or {\bf Lancichinetti} datasets, respectively. 

{\bf II. Real World Datasets.}
In the real world, community structure can be found in nearly any network representing interpersonal relationships.  Some of the most striking group structures are seen in the context of friendship networks, particularly for populations such as adolescents (who are frequently segregated by age, among other factors).  The National Longitudinal Study of Adolescent to Adult Health (referred to as {\bf Add-Health} in the following) is a U.S. national study on adolescents in grades 7-12. It is a large, longitudinal study of students drawn from a sample of American secondary schools, with data being collected during an initial in-school survey followed by four in-home interviews spanning from 1994 to 2008~\cite{ADD_HEALTH}.  In this work we employ the public-use network data from the first wave, in which students were asked to nominate peers they considered to be friends; the data used here comprises 16 different networks, in which each network represents a school (or pair of linked schools in which graduates of the first proceed to the second), each vertex represents a student, and each edge represents a friendship nomination.  For each student, we are also provided with data on gender, race/ethnicity, and grade.  Overall, friendships are strongly segmented by grade, with additional divisions by other demographic characteristics in most schools. 

Finally, we also consider smaller networks frequently encountered when analyzing large-scale network data.  Given a large network and a given node, an \emph{egocentric network} (or \emph{ego net}) for this node is the induced subgraph formed by the node (``ego'')  and its first-order neighborhood (``alters'').  An ego net hence includes all ego-alter ties, together with the alter-alter relationships; cohesive subgroups in the ego net (with ego subtracted) thus mark the \emph{local} communities of alters with whom ego interacts.  Here, we employ the {\bf Facebook}  dataset of Mcauley and Leskovec~\cite{Facebook}, which contains a set of ego-networks derived from the Facebook social network comprising of 10 different networks of size ranging from 52 to 1034 nodes (ego removed for analysis). 

\subsubsection{Baseline Modularity-Targeting Algorithms\label{sec:baselines}}

There are two classes of prior state-of-the-art graph generation algorithms that have been used to explicitly target modularity. We briefly present here these algorithms to facilitate better understanding of our experimental results. 

\noindent {\bf I. \TRAJ Algorithm}
 \TRAJ et al. ~\cite{Trajanovski13} present three different rewiring operations capable of varying the modularity value of a graph with a given partition.  The input to the algorithm is a target modularity value $Q_t^*$, a number of communities $m$ and a total number of links $L$.
In the initialization phase, it creates the communities of nodes in the form of a tree with no more than one link between two communities.
This configuration has the maximum achievable modularity for the given input.
Then, rewiring transformations are applied to lower the modularity value to the target one.

A limitation of this approach is that it assumes the community graph partitioning $c_1^*,\dots,c_n^*$ never changes.  Since the best partition is dictated by the topology, the rewiring process may alter them and, consequently, yield an apparent modularity that is lower than the optimum.
Hence, even if the rewiring process successfully obtains the desired modularity value $Q_t^*$ with respect to the initial partitioning, the output graph may have topology with a slightly different inner community structure, leading to a completely different modularity value $Q_o^*$.  

\noindent {\bf II. Degree Corrected - Stochastic Block Model (DC-SBM)}
The stochastic block model \cite{snijders.nowicki:joc:1997} is a generative algorithm for groups or communities in networks.
The related algorithms work by placing each edge $e_{i,j}$ between the vertices $v_i,v_j$ with a probability which is function of $c_i$ and $c_j$.
Since $c_i,c_j\in \{C_1,\dots,C_m\}$, these probabilities form a matrix of inter and intra community (block) connection probabilities.

Generally, simple stochastic block models have difficulty in building graphs with community structure matching real-world data sets~\cite{dcsbm}.  A limitation of the most basic models in this class is that they do not include effects for other forms of structure found in real-world networks, e.g. degree heterogeneity.
The approach proposed by Karrer and Newman~\cite{dcsbm} rectifies this last limitation by correcting for the degree distribution, yielding improved performance on realistic topologies.


\subsection{Results on Targeting Modularity}
\label{sec:modres}

Here, we evaluate the main objective of this paper, {\em i.e.,} how well \SGF achieves the target modularity.   When comparing the modularity value of an input graph $Q_i$ (or $Q_i^*$) with the modularity value of an output one $Q_o$ (or $Q_o^*$), we consider the (maximum) \emph{modularity ratio} $\frac{Q_o}{Q_i}$ (or $\frac{Q_o^*}{Q_i^*}$): the closer this ratio to 1, the closer the output graph matches the input in terms of modularity.   To compute the modularity value we use the well-known algorithm by Lefebvre et al.~\cite{commdetection}.\footnote{There is a large body of work on community detection algorithms to maximize $Q$~\cite{survey_community_detection}.   In this work, we compute $c_1^*,\dots,c_n^*$ and the associated modularity values using the well-known algorithm by Lefebvre et al.~\cite{commdetection}.}


\subsubsection{Insights into SGF itself.} We first consider a trivial dataset for which our graphs are made of 128 nodes and 2 or 8 communities, and we vary parameters of SGF ($\alpha$) as well as the inherent community structure of the target network to get insights into the behavior of SGF. 
\stdimg{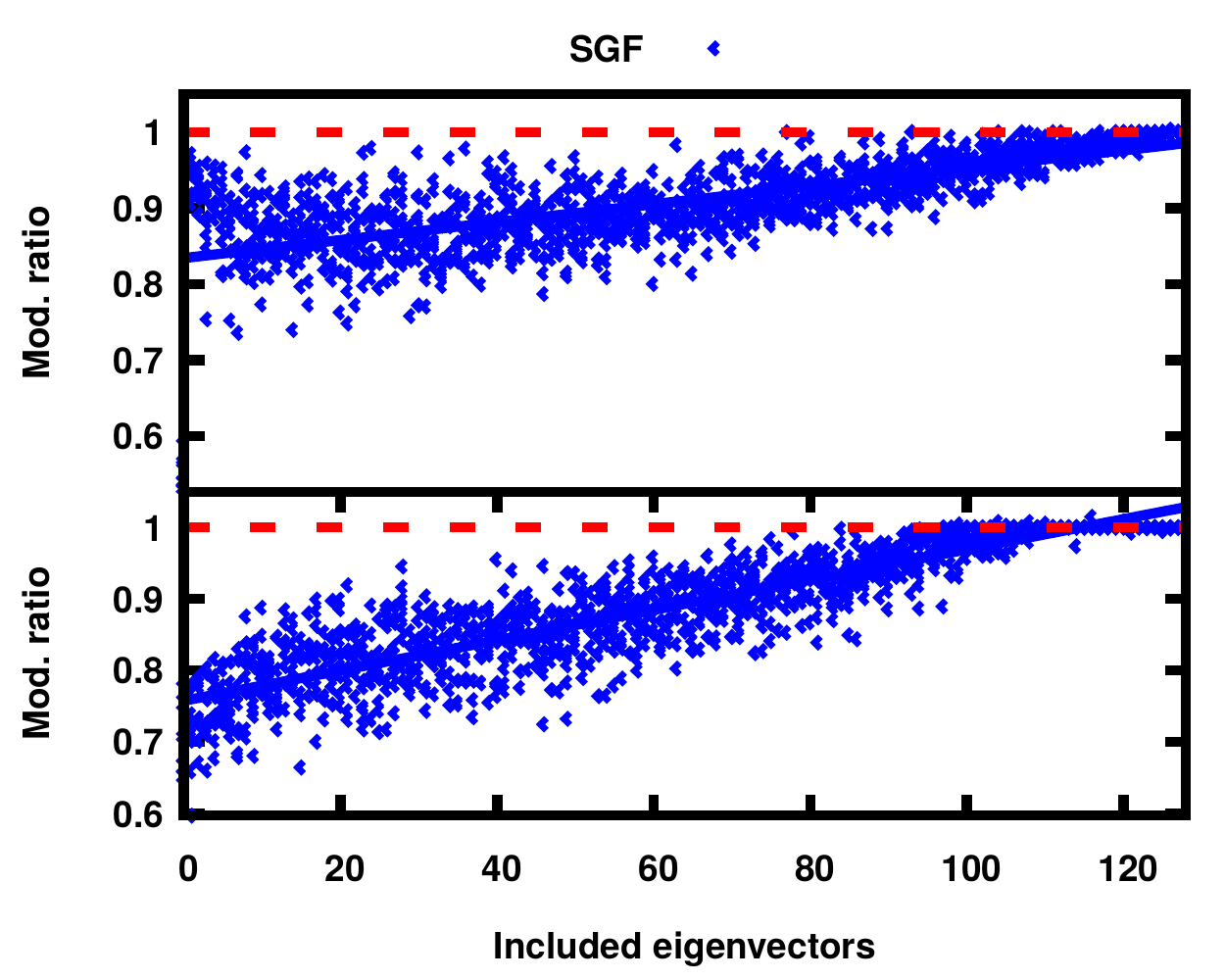}{Modularity ratio for \SGF varying $\alpha$ on our simple graphs with 2 communities (upper plot) and 8 communities (lower plot). }{fig:inc_alpha}
\cref{fig:inc_alpha} shows  the maximum modularity ratio $\frac{Q_o^*}{Q_i^*}$ varying the $\alpha$ parameter with networks of two and eight communities respectively. As expected, the as $\alpha \rightarrow 1$, the more eigenvectors are used and the better the performance  ($\frac{Q_o^*}{Q_i^*} \rightarrow 1$). 

The SGF performance, in terms of how well it targets modularity, is also influenced by the inherent community structure. In \cref{fig:inc_alpha}, we see that SGF requires more eigenvectors for 8 than for 2 communities, in order to obtain similar results.
In \cref{fig:inc_mod}, we see that the higher the modularity of the input graph (controlled by the intra-community edge probability) the better the performance of SGF. 
\stdimg{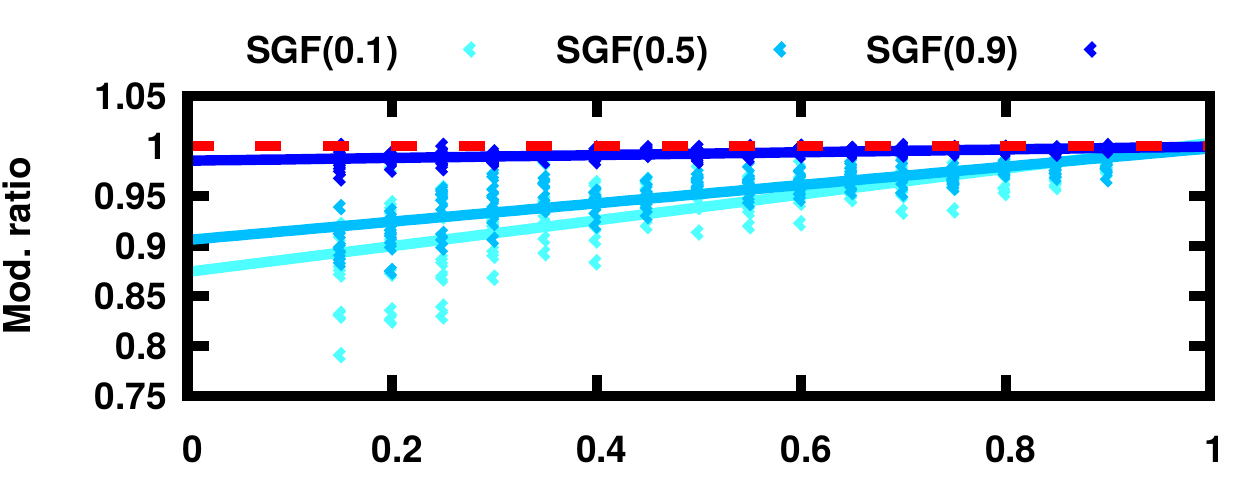}{Modularity ratio for \SGF on our simple graphs varying their connectivity with 2 communities.}{fig:inc_mod}


\subsubsection{Evaluating SGF against baselines.}

{\em Experiments.} In the {\em SGF} pipeline, we target the modularity matrix as the $M=B$, the truncation as the normalization function, and a value $\alpha=0.9$. 
We apply SGF to each input network ten different times, computing the modularity ratio for each simulation; the same process is carried out for the two baseline algorithms.  
In particular, for the Trajanovski et al. algorithm, we compute the modularity value $Q_o^*$, the number of nodes $n$, the number of edges and the number of communities from each input graph and we use them for the Trajanovski approach.
As input to the DC-SBM algorithm, instead, we use the node degree sequence, the node group assignment $c_1^*,\dots,c_n^*$, the group degree sequence on this partition and the number of edges between each pair of groups, computed on the original graph.

{\em Comparison.} \cref{tab:mod_tab} reports the mean and the standard deviation of obtained modularity ratios comparing the methods with different values of $\alpha$ on every dataset and it shows the \SGF requires $\alpha\sim 0.7,0.8$ in order to obtain results similar to the one of DC-SBM and that Trajanovski obtains results with a very high standard deviation from the target value.
\cref{fig:modratio} reports the mean and 99\% confidence intervals for the resulting modularity ratios, for which SGF with $\alpha=0.9$ consistently produces graphs with a maximum modularity nearly equal to the input data set (modularity ratio appx. 1). In contrast, the Trajanovski et al. algorithm frequently produces graphs with too high or too low modularity, and does so with high variance.  The DC-SBM performs more consistently, but tends to achieve modularity lower than  the target, and generally shows higher variance.  Interestingly, both the Trajanovski et al. and DC-SBM algorithms perform better on the synthetic data sets than in their real-world counterparts.  Across all four data sets, the \SGF is consistently closer to the target.
\stdimg{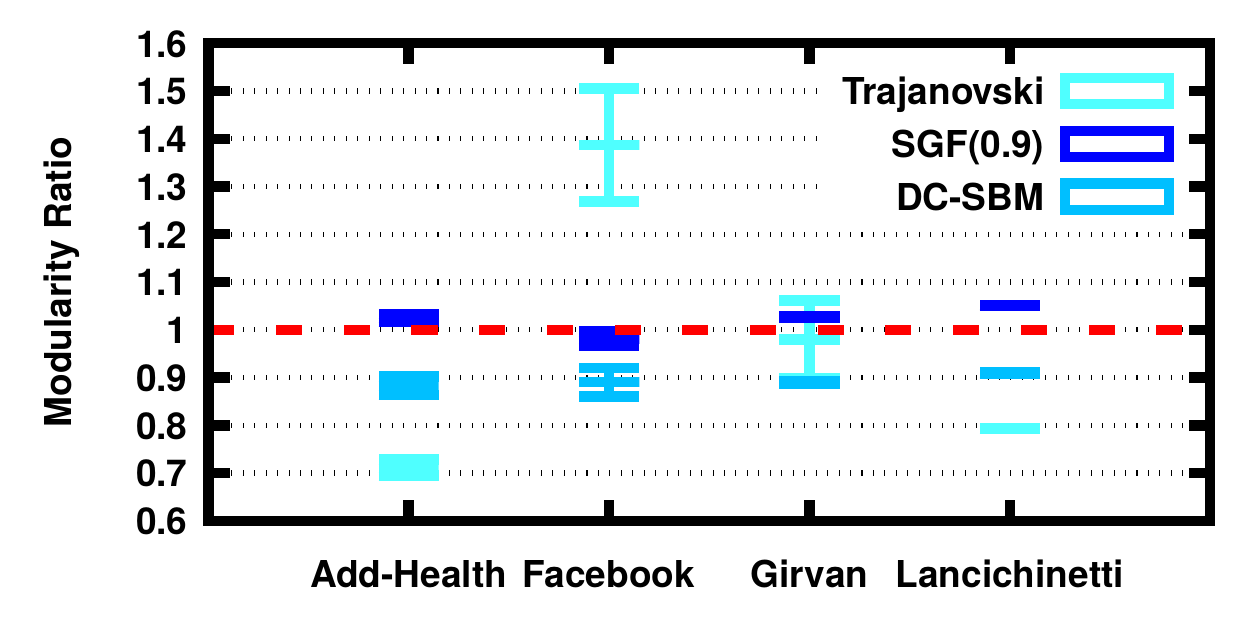}{Means and 99\% confidence intervals for the modularity ratio, by method and data set.}{fig:modratio}

\begin{table}
	\caption{Modularity ratio for all the strategies on all the datasets. }
	\label{tab:mod_tab}
	\centering
	{\tiny
	\begin{tabular}{|l||lcc|lcc|}
\hline
Strategy & Dataset & Mean & Std & Dataset & Mean & Std\\
\hline
\hline
SGF(0.1) &  Add-Health & 0.76038 & 0.10479 &  Facebook & 0.80551 & 0.07143\\
SGF(0.1) &  Girvan & 0.74602 & 0.01709 &  Lancichinetti & 0.58268 & 0.00799\\
\hline
SGF(0.2) &  Add-Health & 0.71985 & 0.06740 &  Facebook & 0.82773 & 0.08925\\
SGF(0.2) &  Girvan & 0.70728 & 0.01033 &  Lancichinetti & 0.54349 & 0.00576\\
\hline
SGF(0.3) &  Add-Health & 0.75379 & 0.07496 &  Facebook & 0.87341 & 0.08013\\
SGF(0.3) &  Girvan & 0.70690 & 0.01553 &  Lancichinetti & 0.55989 & 0.00530\\
\hline
SGF(0.4) &  Add-Health & 0.80250 & 0.10278 &  Facebook & 0.86587 & 0.06847\\
SGF(0.4) &  Girvan & 0.73735 & 0.00793 &  Lancichinetti & 0.59954 & 0.00445\\
\hline
SGF(0.5) &  Add-Health & 0.82452 & 0.08711 &  Facebook & 0.89184 & 0.06929\\
SGF(0.5) &  Girvan & 0.77007 & 0.01134 &  Lancichinetti & 0.64635 & 0.00339\\
\hline
SGF(0.6) &  Add-Health & 0.86057 & 0.08472 &  Facebook & 0.91149 & 0.06688\\
SGF(0.6) &  Girvan & 0.83302 & 0.01151 &  Lancichinetti & 0.71476 & 0.00564\\
\hline
SGF(0.7) &  Add-Health & 0.88024 & 0.05368 &  Facebook & 0.93420 & 0.07223\\
SGF(0.7) &  Girvan & 0.88050 & 0.01007 &  Lancichinetti & 0.78831 & 0.00714\\
\hline
SGF(0.8) &  Add-Health & 0.93563 & 0.03820 &  Facebook & 0.96249 & 0.02018\\
SGF(0.8) &  Girvan & 0.92902 & 0.00797 &  Lancichinetti & 0.87126 & 0.00553\\
\hline
SGF(0.9) &  Add-Health & 1.02524 & 0.03708 &  Facebook & 0.98131 & 0.05789\\
SGF(0.9) &  Girvan & 1.02722 & 0.00429 &  Lancichinetti & 1.05135 & 0.00338\\
\hline
DC-SBM &  Add-Health & 0.88304 & 0.09229 &  Facebook & 0.89041 & 0.11375\\
DC-SBM &  Girvan & 0.88989 & 0.00892 &  Lancichinetti & 0.90944 & 0.00462\\
\hline
Trajanovski &  Add-Health & 0.71150 & 0.08293 &  Facebook & 1.38715 & 0.45120\\
Trajanovski &  Girvan & 0.97946 & 0.31149 &  Lancichinetti & 0.79338 & 0.00097\\
\hline
\end{tabular}

	}
\end{table}

\stdimg{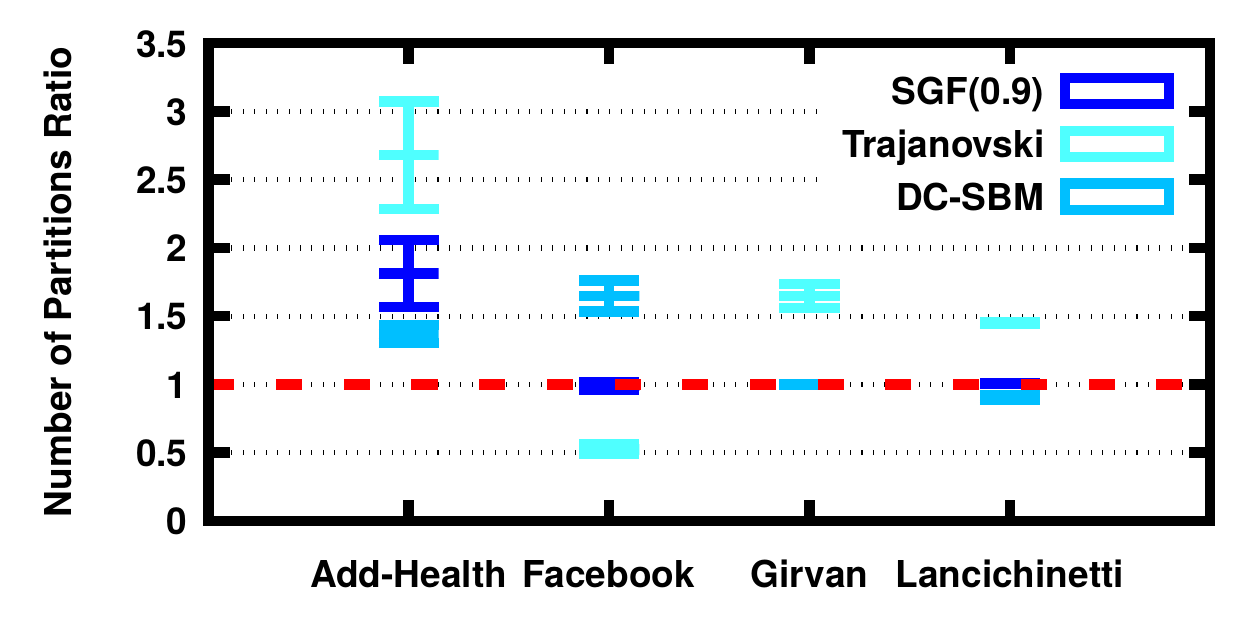}{Means and 99\% confidence intervals for the partition number ratio, by data set. Note that the SGF$(0.9)$ and DC-SBM algorithm results overlap with the Girvan data set graphs.}{fig:modnumratio}

\cref{fig:modnumratio} looks at a related but different metric: it presents means and 99\% confidence intervals for the ratio of {\em number of detected output partitions} (under modularity maximization) vs. the number of input partitions; this expresses the extent to which each algorithm preserves the number of communities found in the input graph. 
The target value is again 1, as our objective is to generate graphs with the same number of communities.  \SGF consistently reproduces the number of input partitions for three out of the four data sets, but tends to split communities within AddHealth.  None of the other algorithms consistently performs as well as the \SGF, with some being tied on some data sets; the only case of better performance by a competing algorithm is DC-SBM on AddHealth, which overproduces communities by a slightly smaller margin.\footnote{ None of the three algorithms perform optimally on this data set. That can be due to input partition solutions not necessarily stable to small perturbations, and a generated graph could have different numbers of identified subgroups than the source graph even if the two are similar in most respects.}

\subsection{Results on Other (Local) Structural Properties}
\label{sec:other_metrics}
In addition to accurately targeting modularity, 
our approach also generates output graphs $A'=SGF(\alpha)$ maintaining other important (local structural) properties of the input $A$.

\stdimg{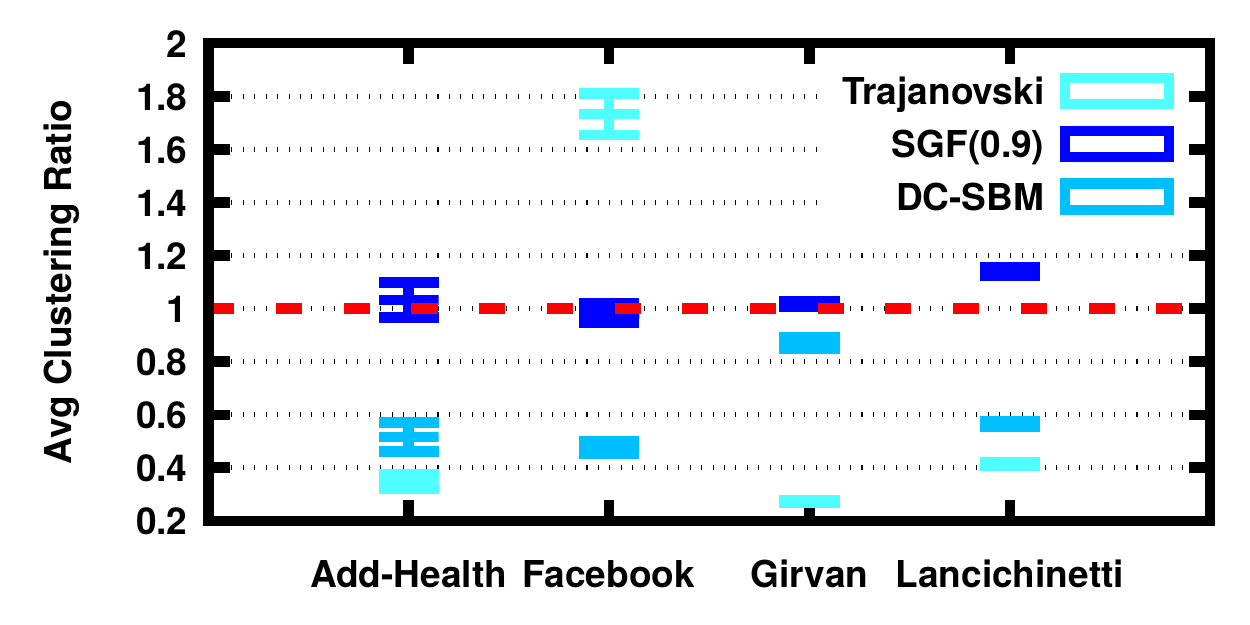}{Means and 99\% confidence intervals for average clustering ratio, by data set.}{fig:clust}
{\em Clustering} is a local network property, indicating the extent of triadic closure. \cref{fig:clust} presents means and confidence intervals for the average clustering ratio between the output  and input graphs.  
Although it is not apparent that the \SGF would do well at capturing this local property - since it targets global structure - it in fact does well for all four data sets.  By contrast, the other modularity-targeting methods do not tend to preserve clustering.  This is particularly evident for the Trajonovski et al. algorithm, whose rewiring strategy can easily alter the structure of triads.

\stdimg{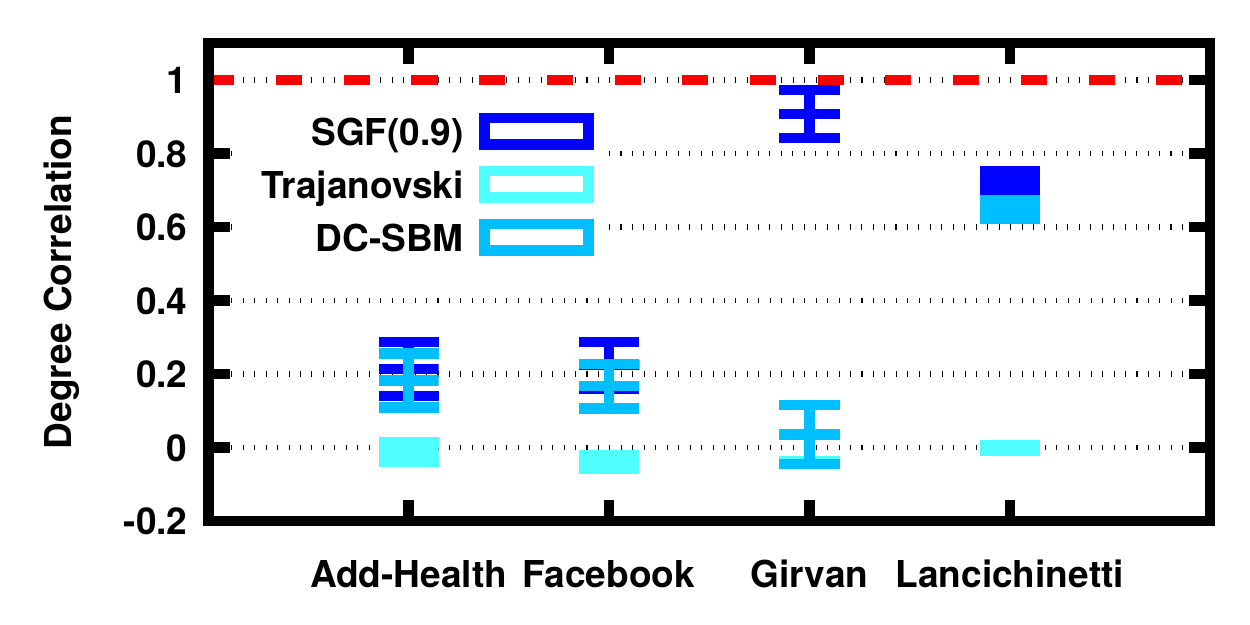}{Means and 99\% confidence intervals for degree sequence correlation, by data set. Note that the DC-SBM result lower bound overlaps the \TRAJ results with the Girvan dataset.}{fig:degree}
{\em Degree sequence.}  We also examined the correlation of the degree sequence of the input and output graphs.  (Recall that, in the presence of covariates, nodes are non-exchangeable.)  While the eigenvectors of the modularity matrix are only loosely connected with degree, we show better performance on this metric than the competing methods.  In particular, \TRAJ doesn't attempt to preserve the degree distribution at all, leading (reasonably enough) to a nearly complete loss of correlation between degree sequences.  The relatively poor performance of the DC-SBM method is more surprising, given that it attempts to preserve degree information.  Overall, it performs only slightly worse than the \SGF for three out of four data sets, the lone exception being the Girvan synthetic networks (where the \SGF has a correlation of nearly 1, vs. almost 0 for DC-SBM).

\subsection{Results on Attribute Modularity Preservation}
\label{sec:attributes}
We find that our approach not only successfully preserves the maximum modularity value $Q^*$ but also preserves the modularity of significant partitions other than the topologically best one.
If we consider partitions $c_1,\dots,c_n$ indicating e.g. gender, race/ethnicity, or school grade provided by the Add-Health dataset, we can compute the resulting modularity values $Q_i$, on the original graphs, and $Q_o$ on the output graphs after applying the same attribute sequence, summarizing the similarity by the resulting modularity ratio. 
The closer this ratio  is to 1, the better maintained the community structure for partitions associated with important covariates (whether or not they correspond to modularity-maximizing partitions).

\cref{fig:modattr} shows the comparison. The Add-Health dataset is a real-world dataset with attributes associated with the nodes.
Attribute labeling for the output graphs of \TRAJ and DC-SBM is done trivially, by assigning attributes to the nodes in the same order as the input ones.  As expected, the \SGF closely preserves modularity on all attribute partitions, while the other methods typically perform poorly, entirely losing the graph structure.
This is not surprising as the other methods were not designed to preserve other community structure but the topological one and their performance are shown only for completeness.
On contrary, SGF successfully reproduce all the possible (even overlapping) community structures from the original ones, being, to the best of our knowledge, the first suitable generator for creating graph proxies with complex community structures.

\stdimg{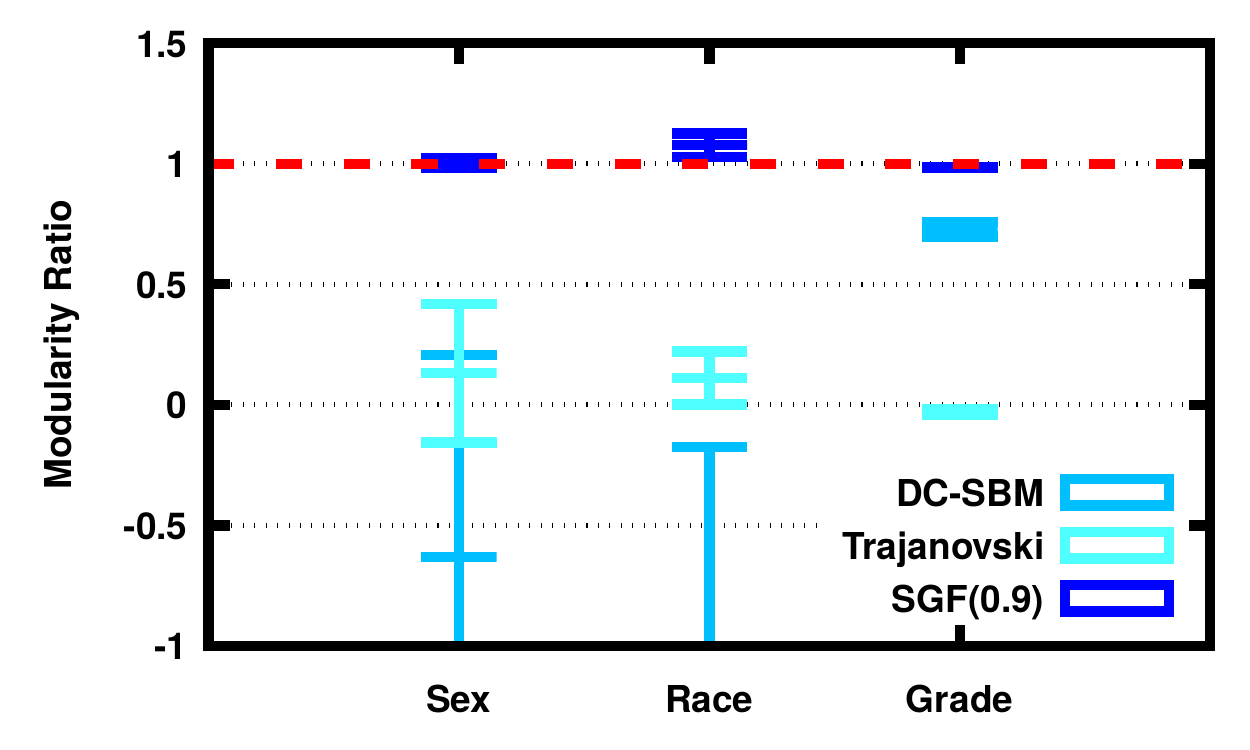}{99\% confidence interval on attribute modularity ratio for the Add-Health dataset.}{fig:modattr}

\subsection{Randomness of Realizations and (De)anonymization}
\label{sec:deanonym}
The closer $\alpha$ is to 1, the better our approach targets the intended modularity.
However, two natural questions arise from this relationship: (1) as $\alpha \rightarrow 1$, how concentrated is the distribution of graphs $A'=SGF(\alpha)$ we produce? (2) as $\alpha \rightarrow 1$, how distinct is our output $A'$ from the input $A$?
Ideally, we would like a relatively low level of concentration and a reasonable distance from the input graph, to grant a sufficient degree of randomness in our realizations.  We use the normalized Shannon entropy to measure the concentration of the SGF distribution in the space of graphs; this is important for the use case of simulation.
To measure the distinctness of a realized graph from the input network, we consider a pragmatic metric that maps to another important potential use case: {\em resistance to  de-anonymization attacks}. We envision that \SGF could be used not only as a graph generator for simulating graphs resembling real-world networks, but also as an anonymization technique (in conjunction with others) to anonymize $A$ to $A'$, while preserving several global and local properties.

De-anonymization  attacks attempt to identify the nodes in a partially labeled graph, 
 exploiting similarities between the two graphs and potentially auxiliary informartion. To the extent that a generated network cannot be readily de-anonymized with respect to the original graph structure, the two networks are clearly distinct; this is important for applications such as anonymization of real network datasets,  where concealment of node identities is a major priority.  For example, a Facebook dataset may be collected  and made publicly available after proper anonymization (removing node ids, and perturbing the network structrure); de-anonymization attacks can infer the node ids based on graph structure and auxiliary information.  
  A survey of state of the art anonymization and de-anonymization techniques can be found in \cite{secgraph}.
We pick one of the state-of-the-art de-anonymization attacks: the Distance Vector attack~\cite{DV}, which is proven to be scalable, robust and exploitative of global graph characteristics~\cite{secgraph} preserved by the SGF.  At each run, we feed the Distance Vector attack with a seed of 5\% of nodes as ground truth and we report the mean fraction of nodes the attack successfully identified.  Examining the fraction of nodes identified as a function of $\alpha$ allows us to examine the tradeoff between privacy preservation and synthetic data quality. 

\stdimg{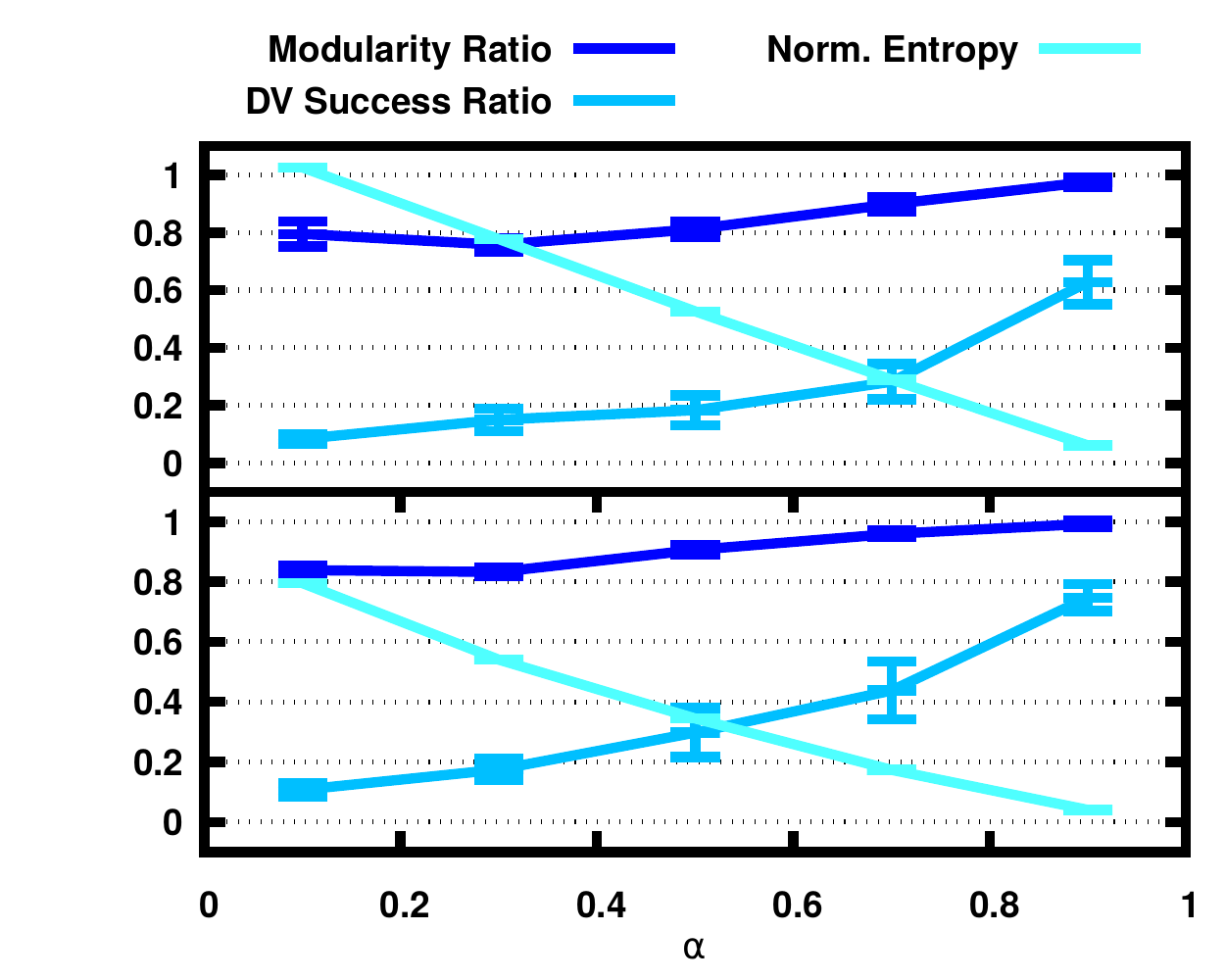}{Means and 99\% confidence intervals for graph distribution entropy and DV de-anonymization success rates on one Add-Health graph (upper plot) and one Facebook graph (lower plot), by $\alpha$.}{fig:deanonym}

The upper plot of \cref{fig:deanonym} presents the results for one connected graph from the Add-Health dataset.
We note that the modularity ratio remains close to the target 1 even for low values of $\alpha$, implying that it is possible to preserve much of the community structure with a fairly small fraction of eigenvectors.  Entropy is close to 0 for very large $\alpha$, but climbs steadily as $\alpha$ is reduced, indicating the extent of reduction in concentration.  More striking is the success rate of the DV attack.  Even at $\alpha=0.9$, the attack can identify only around 60\% of nodes from the original graph given the synthetic graph and a 5\% seed sample.  
This falls apace, with a success rate of only around 20\% once $\alpha=0.5$ (corresponding to a modularity ratio $\sim0.82$ from \cref{tab:mod_tab}), and a rate not significantly different from baseline once $\alpha=0.25$ (modularity ratio $\in (0.71,0.75)$).  Given the strength of this attack, this implies a high degree of discrepancy between the initial and target graph (while maximum modularity is preserved). Similar results are reported in the lower plot of \cref{fig:deanonym} for a connected graph from the Facebook dataset.
Again, even for high values of $\alpha$ the de-anonymization attack cannot fully align the graphs.

For both datasets, when $\alpha=0.1$, the  modularity ratio is still close to the target 1 and the de-anonymization attack can identify only approximately 10\% of nodes (while knowing 5\% in the first place).
This suggests that the SGF has the potential to be used as a tool (in combination with others \cite{secgraph}) for anonymizing sensitive graph datasets.


\section{Conclusion}
\label{sec:conclusion}

We propose \SGF (SGF) - a framework for generating graphs that resemble real-world networks in terms of global properties, via spectral decomposition and approximation. Our focus in this paper was the use of SGF specifically to generate graphs with a target modularity; we showed that  it succeeds in doing so and it outperforms baselines. At the same time, SGF preserves other important local structure properties (such as clustering and the degree sequence) and node attribute structure. 
It also does well at preserving modularity associated with such node attributes, even when they are not specifically targeted;
this allows SGF to generate reliable graph proxies for real-world networks with arbitrarily complex and overlapping community structures.
The parameter $\alpha$ can control not only the error in modularity ($\alpha \rightarrow 1$ brings us arbitrarily close the target modularity) but also  the entropy of graph distribution. Lowering $\alpha$ leads to higher entropy  of the realizations, thus making  the  node identities in the input graph increasingly difficult to reveal via de-anonymization attacks.  This makes  SGF a potentially useful tool (in combination with existing anonymization techniques) for generating synthetic data sets from sensitive data.
%
Our codebase is available on-line~\cite{codebase}.


\bibliographystyle{IEEEtran}
\bibliography{biblio}

\newpage

\section*{Appendix A: $\alpha$-Tunable Convergence of $\tilde{M}$ to $M$}
The distance between $M$ and $\tilde{M}$ due to the low pass filter at the second step of the pipeline indicates how much information we dropped from the transformed matrix $M$.
We measure this distance using the Euclidean norm, $\left\lVert M - \tilde{M} \right\rVert_2$, a well-known metric on spaces of real-valued matrices.
This is distance is directly related to the spectral radius:
$$\left\lVert M - \tilde{M} \right\rVert_2 = \left\lVert \sum_{i=\lceil\alpha n\rceil+1}^{n} \lambda_i v_i v_i^T \right\rVert_2 = \rho \left(\sum_{i=\lceil\alpha n\rceil+1}^{n} \lambda_i v_i v_i^T\right),$$
where $\rho(\cdot)$ is the spectral radius function.
Knowing that the eigenvalues of $M$ are ordered with respect their moduli, $|\lambda_1|\geq \dots \geq |\lambda_n|$, we obtain
$$\left\lVert M - \tilde{M} \right\rVert_2 =\lambda_{\lceil\alpha n\rceil+1},$$
which is the largest eigenvalue not included in our approximation $\tilde{M}$.
Hence, our method allows us to target the reproduction of $M$ with arbitrary precision in the Euclidean norm.  Indeed, we can guarantee exact preservation of $M$ in the limit as $\alpha \to 1$, since $\forall \epsilon \in \mathbb{R}^+\ \exists \alpha \in [0,1] :\left\lVert M - \tilde{M} \right\rVert_2 \leq \epsilon$.  Typically, we will deliberately select $\alpha<1$ to remove information in $M$ that is idiosyncratic to $A$, but the \SGF allows users to bring $\tilde{M}$ as close to $M$ as needed for the application in question.

\section*{Appendix B: Impact of Normalization on $M$-preservation and Entropy}
\label{sec:test_entropy}
\label{sec:convergence}
\label{sec:entropy}

\stdimg{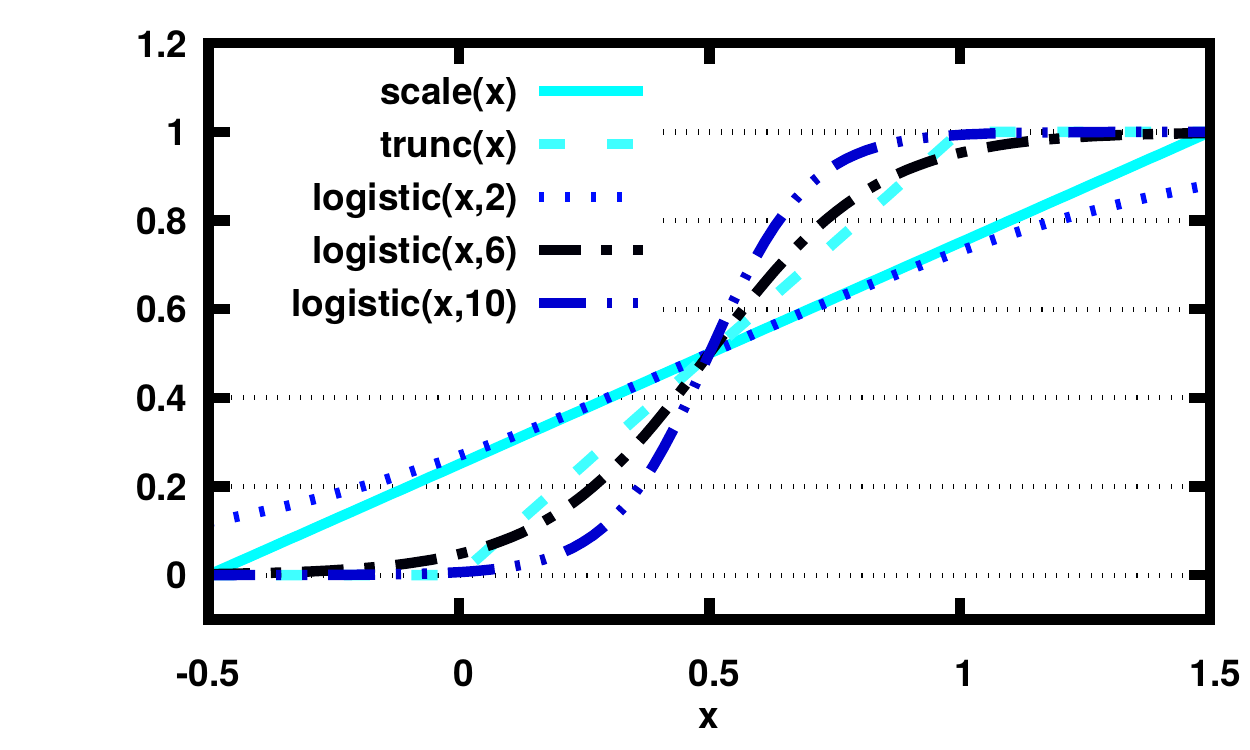}{Normalization function comparison in the interval $(0.5,1.5)$.}{fig:norm_function}

\cref{fig:norm_function} presents the different normalizing functions in the interval $(-0.5,1.5)$.
The normalization applied by \textit{scale()} introduces a distortion highly dependent on the difference between the minimum and maximum values (with $k=6$ providing the best performance) and  the \textit{logistic()} introduces noise by remaining strictly within $(0,1)$ for all $\tilde{A}_{i,j}$.
Preliminary experiments showed that with $k=6$ the logistic function provided the best performance in the cases studied here, and we use this value in the analyses that follow.

In addition to spectral filtering, details of the original input graph are removed by normalization. It is hence useful to examine how different choices of normalization function alter both the degree of ``smoothing'' in the graph generation process (measured by the mean Euclidean norm between the initial and generated graph) and the entropy of the resulting graph distribution.  To this end we measure the distortion low-rank approximation introduces in an adjacency matrix through a simulation experiment with input graphs drawn from a set of realizations from \ER and \BA graph generation processes; our input set was composed of 10 random \ER graphs and 10 random \BA graphs, each having $n=100$ nodes and the same approximate mean degree of $\sim 4.5$. 


\stdimg{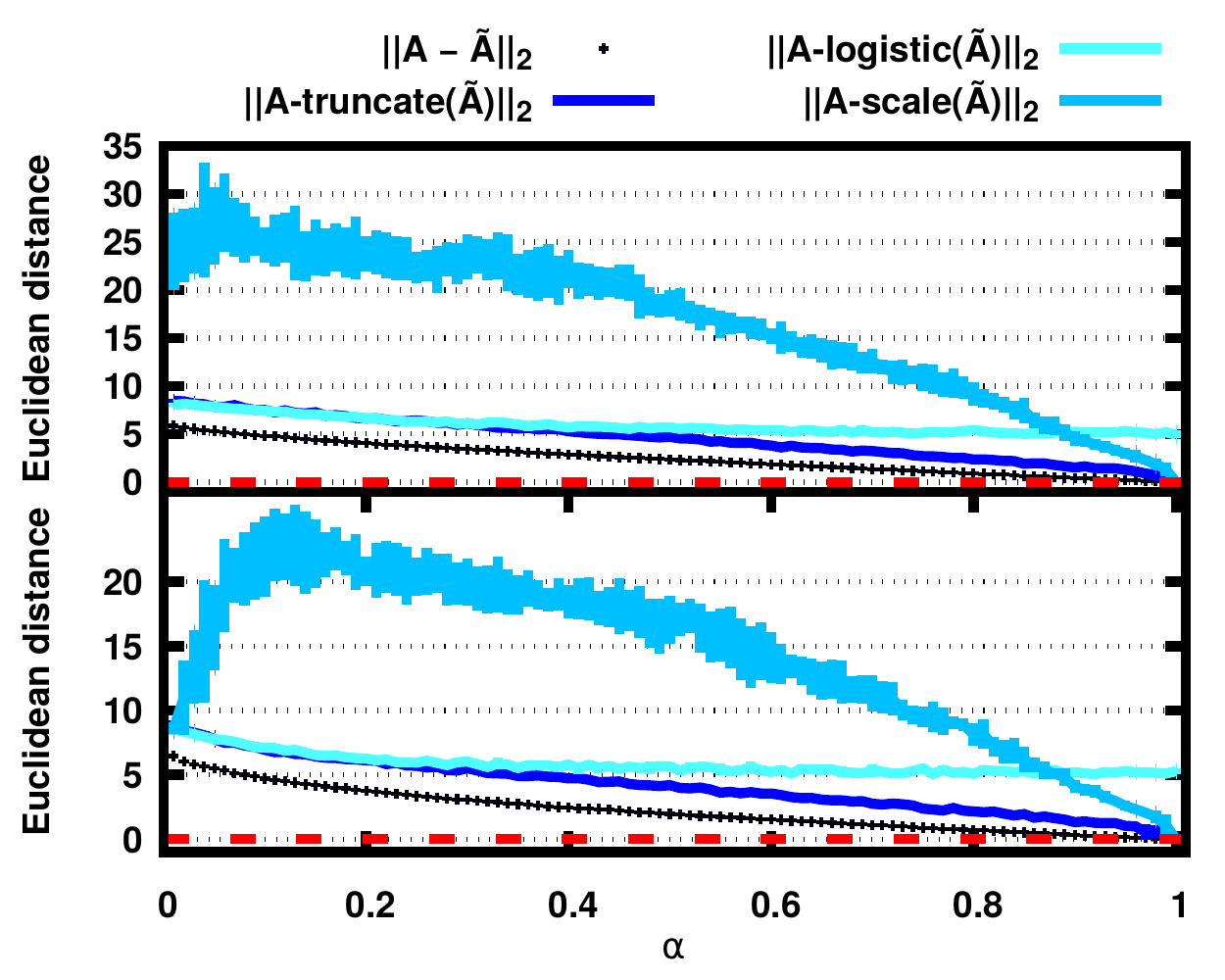}{Baseline and normalized $A,\tilde{A}$ distances for \ER graphs (upper plot) and \BA graphs (lower plot), by $\alpha$. Ideally, normed matrix distances should be close to $\lVert A - \tilde{A}\rVert$. }{fig:eigerr}

\cref{fig:eigerr} shows Euclidean norms for the respective samples, as a function of $\alpha$.  
These norms represent matrix distances and, hence, they span in the interval $(0,\infty)$.
The patterns are quite similar for \ER and \BA graph types, with the exception of the limiting distance in the case when $\alpha \to 0$ (where at most one eigenvalue is employed).  As noted in section~\ref{sec:convergence}, $\lVert A - \tilde{A} \rVert \to 0$ as $\alpha \to 1$ (since we have chosen $M=A$); however, some of this information may be lost by normalization.  
In particular \cref{fig:eigerr} shows a comparison among the normalization functions and the distance they introduce from the spectral-approximated matrix $\tilde{A}$.
Ideally, we do not want the normalization function to introduce systematic variation in this space, as the approximation introduced by SGF should rely only on the spectral approximation.
Hence, the distance $\parallel A-\text{norm}(\tilde{A})\parallel$ should be as close as possible to $\parallel A - \tilde{A}\parallel$.
As can be noted in \cref{fig:norm_function} the scale function introduces considerable noise, a phenomenon also seen in \cref{fig:eigerr}.

Both truncation and logistic approaches obtain a distance very close to the spectral approximation, but only truncation preserves convergence to $0$ when $\alpha\to 1$.

\cref{fig:entropy} shows the entropy computation for the same graphs during the same runs.  
Entropy here represents the expected number of bits required to describe $A'$ given the generating distribution. We normalize this value for a fair comparison among graphs with different densities.
As would be expected, entropy is lowest when all eigenvectors of the original graph are included, and increases as constraints are reduced.  However, the choice of normalization function affects the extent to which information from $\tilde{A}$ contributes to $A'$.  Overall, scaling and truncation rules show the smoothest entropy enhancement, with the logistic rule showing the least effect; this is because the prior functions yield a larger change in tie probability given a fixed change in cell value over the range encountered in these tests.  
It is worth noticing the highly variable behaviour of the \textit{scale()} function which, although it follows a clear trend, it is strongly influenced by the maximum and minimum values of $\tilde{A}$, leading to high variance and non-monotonicity when $\alpha \to 0$.

\stdimg{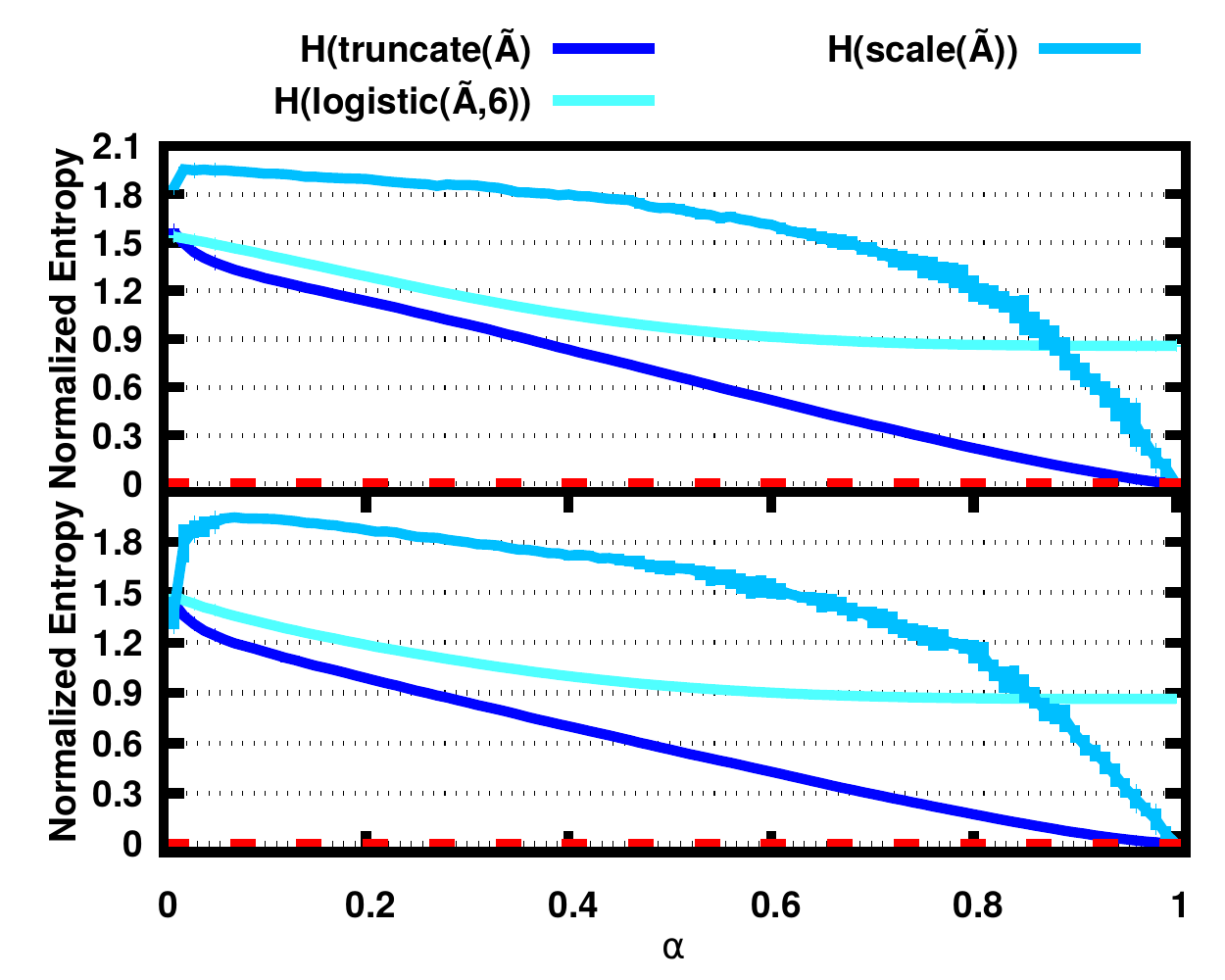}{Entropy of generated graphs using three different normalization functions for \ER input graphs (upper plot) and \BA graphs (lower plot) computed for different values of $\alpha$.}{fig:entropy}

The above suggest that the cleanest entropy and distance relationships are found for the truncation rule.

\end{document}